\documentclass{jps-cp}
\usepackage{txfonts} 
\usepackage{bm}
\usepackage{graphicx}
\usepackage{color}

\title{BCS-BEC Crossover of Triplet Exciton Condensation in Bilayer Systems}

\author{Hidemaro \textsc{Suwa}$^{1}$, Shang-Shun \textsc{Zhang}$^{2}$, and Cristian D. \textsc{Batista}$^{2}$}

\inst{$^{1}$Department of Physics, University of Tokyo, Tokyo 113-0033, Japan\\
$^{2}$Department of Physics and Astronomy, University of Tennessee, Knoxville, TN 37996, USA
}

\email{suwamaro@phys.s.u-tokyo.ac.jp}

\recdate{July 31, 2022}

\abst{
We study the BCS-BEC crossover phenomenon of triplet exciton condensation using a half-filled bilayer Hubbard model.
We calculate the dynamical spin structure factor and the exciton wave function on the phase boundary between the antiferromagnetically ordered and disordered phases.
In the BCS regime, the formation and condensation of particle-hole bound states, namely excitons, occur simultaneously at the phase transition point, and the exciton wave function is extended in real space.
In the BEC regime, on the other hand, bound states are well defined over the whole Brillouin zone also in the disordered phase, and the size of  particle-hole pairs is smaller than  a lattice space.
Quantum criticality of charge-spin-orbital entangled states can emerge from intermediate coupling materials in the crossover regime.
}

\kword{exciton condensation, BCS-BEC crossover, excitonic insulator, spin-orbit coupling, antiferromagnet, bilayer, iridate}

\begin{document}
\maketitle

\section{Introduction}
\vspace{-2mm}
Quantum criticality is one of the central topics in condensed matter physics~\cite{Sachdev2011}.
A typical physical system showing quantum criticality is the bilayer antiferromagnet in the strong coupling limit, in which the quantum phase transition occurs at a certain ratio of intralayer and interlayer exchange couplings~\cite{Lohofer2015}.
In the strong interlayer hopping limit, the ground state of the magnetic system is the direct product state of singlet dimers on each vertical bond between the two layers.
The low-energy excitation is represented by triplons, and the quantum phase transition to the magnetically ordered phase is described as a  triplon condensation.
In the strong coupling regime, the electron-hole binding occurs at an energy scale of the order of the charge gap, and the triplon condensation happens at a much lower magnetic energy scale.
This condensation mechanism is characteristic of the BEC regime.
The triplon state is adiabatically connected to the exciton state that arises in the weak coupling regime, in which the electron-hole binding and the exciton condensation occur simultaneously. This is the so-called  BCS regime.
The BCS-BEC crossover has been extensively studied in the superfluid phase transition of Fermi gases in optical lattices~\cite{Ohashi2020}.
The interaction between atoms, or the scattering length in the $s$-wave channel, can be experimentally controlled by the Feshbach resonance.
While largely overlapping one another in the BCS regime, Cooper pairs shrink and behave like molecular bosons in the BEC regime.
In the meantime, a variety of spin-orbit insulators have been recently found in $5d$-orbital electron systems, such as iridates~\cite{Rau2016,Cao2018}.
In these materials, the charge gap is comparable to the magnetic energy scale, indicating that the system is in the intermediate coupling, namely the BCS-BEC crossover regime.
Enhanced entanglement in the crossover regime between charge, spin, and orbital degrees of freedom provides rich many-body physics and functional quantum materials~\cite{Hao2018}.

Exciton condensation induces an excitonic insulator, which is classified into four types according to the spin of the excitons and the phase of the order parameter~\cite{Jerome1967,Halperin1968}.
The condensation of a soft charge mode, namely singlet excitons, has been discussed in TmSe$_{0.45}$Te$_{0.55}$~\cite{Bucher1991}, 1{\it T}–TiSe$_2$~\cite{Cercellier2007}, Ta$_2$NiSe$_5$~\cite{Wakisaka2009}, etc.
Recently, a spin-triplet exciton condensation was evidenced by resonant inelastic X-ray scattering of the bilayer iridate Sr$_3$Ir$_2$O$_7$, considered in the crossover regime~\cite{Mazzone2022}.
It is then important to study the BCS-BEC crossover phenomenon~\cite{Phan2010,Seki2011} of exciton condensation in the presence of multiple entangled degrees of freedom and to understand how triplet exciton condensation in the crossover regime affects relevant quantities and measurements.

In the present paper, we study the crossover behavior of the quantum critical point (QCP) using a single-band bilayer Hubbard model, by extending the argument of the exciton condensation given in Ref.~\cite{Suwa2021}.
By calculating the dynamical spin structure factor and the exciton wave function in the mean field and the random phase approximation (RPA), we reveal how the BCS-BEC crossover phenomenon takes place in the bilayer system.

\section{Model}
\label{sec:model}
\vspace{-2mm}
We study a half-filled bilayer Hubbard Hamiltonian that models bilayer systems of the $d^5$ electron configuration, such as Sr$_3$Ir$_2$O$_7$~\cite{Carter2013}.
In this material, the $t_{2g}$ orbitals of the Ir ions are split into the $J_{\rm eff}=1/2$ and $3/2$ orbitals by spin-orbit coupling.
The low-energy physics can be described by the following single-band Hubbard model:
$\mathcal{H}=-\mathcal{H}_{{\rm K}}+\mathcal{H}_{{\rm I}}$,
with $\mathcal{H}_{{\rm I}}\!=U\sum_{\bm{r}}n_{\bm{r}{\uparrow}}n_{\bm{r}{\downarrow}}$
and
\begin{equation}
\!\!\!\mathcal{H}_{{\rm K}}\!=\sum_{{\bm{r}},\bm{\delta}_{\nu}}t_{\nu}{\bm{c}}_{\bm{r}}^{\dagger}{\bm{c}}_{{\bm{r}}+{\bm{\delta}}_{\nu}}\!\!+t_{z}\!\!\sum_{{\bm{r}}_{\bot}}\!{\bm{c}}_{({\bm{r}}_{\bot},1)}^{\dagger}e^{i\frac{\alpha}{2}\epsilon_{\bm{r}}\sigma_{z}}{\bm{c}}_{({\bm{r}}_{\bot},2)}\!\!+\!{\rm H.c.},
\label{eq:Hk}
\end{equation}
where $t_\nu \in\mathbb{R}$ ($\nu$ = 1, 2) are the nearest- and next-nearest-neighbor hopping amplitudes within the square lattice of each Ir layer, 
$t_z \in\mathbb{R}$ is the hopping amplitude on the vertical bonds between the two layers, and $\sigma_z$ is the Pauli matrix. 
The overall phase was chosen to gauge away the phase of $t_\nu$. 
The operator $ c_{\bm{r}}^{\dagger}$ = [$ c_{\uparrow,\bm{r}}^{\dagger}$, $c_{\downarrow,\bm{r}}^{\dagger}$] creates the Nambu spinor of the electron field at $\bm{r} = (\bm{r}_\bot, l)$ with $l = 1, 2$ denoting the layer index and $\bm{r}_\bot = r_1\bm{a}_1 + r_2\bm{a}_2$. 
Here, the primitive in-plane lattice vectors are denoted by $\bm{a}_1$ and $\bm{a}_2$, and the directed neighboring bonds are represented by $\bm{\delta}_1=\bm{a}_1,\bm{a}_2$ and $\bm{\delta}_2=\bm{a}_1 \pm \bm{a}_2$. 
In the interaction term $H_{\rm I}$. $U$ is the effective on-site Coulomb interaction, and $n_{\bm{r}\sigma}$ is the number operator of spin-$\sigma$ electrons at $\bm{r}$.
The sign $\epsilon_{\bm{r}}$ of the spin-dependent hopping term takes the values $\pm1$ depending on which sublattice of the bipartite bilayer system $\bm{r}$ points to. 
The key phase $\alpha$ arises from hopping matrix elements between $d_{xz}$ and $d_{yz}$ orbitals, which are allowed to be nonzero by the staggered octahedral rotation in the unit cell~\cite{Carter2013}.
The system has an easy
$z$-axis spin anisotropy for $\alpha\neq0$, and the ground state
can have Néel ordering, $\langle S_{\bm{r}}^{\mu}\rangle=(-1)^{\gamma_{\bm{r}}}M\delta_{\mu z}$,
where $\gamma_{\bm{r}}=(1+\epsilon_{\bm{r}})/2$, ${S}_{\bm{r}}^{\mu}={1/2}c_{\bm{r}}^{\dagger}\sigma^{\mu}c_{\bm{r}}$
$(\mu=x,y,z)$ and $M$ is the magnetization. 

Following Ref.~\cite{Suwa2021}, we diagonalize the Hamiltonian in the mean-field (Hartree-Fock) approximation and obtain the eigenvalues and the eigenfunctions.
The magnetic susceptibilities of the transverse
and the longitudinal modes are calculated using the RPA
\begin{equation}
\chi^{+-}(\bm{q},i\omega_{n}) = \frac{1}{\tau^{0}-U\chi_{0}^{+-}(\bm{q},i\omega_{n})}\chi_{0}^{+-}(\bm{q},i\omega_{n}), \qquad
\chi^{zz}(\bm{q},i\omega_{n}) = \frac{1}{\tau^{0}-\frac{U}{2}\chi_{0}^{zz}(\bm{q},i\omega_{n})}\chi_{0}^{zz}(\bm{q},i\omega_{n}),\label{eq:rpa_L}
\end{equation}
respectively, where $\tau_{0}$ is the $2\times2$ identity matrix.
Here $\chi^{+-}$ and $\chi^{zz}$ are $2\times2$ matrices in the
sublattice space, while $\chi_{0}^{+-}$ and $\chi_{0}^{zz}$ are
the bare magnetic susceptibilities.

\begin{figure}
    \centering
    \includegraphics[bb=0 0 2216 932, width=0.95\columnwidth]{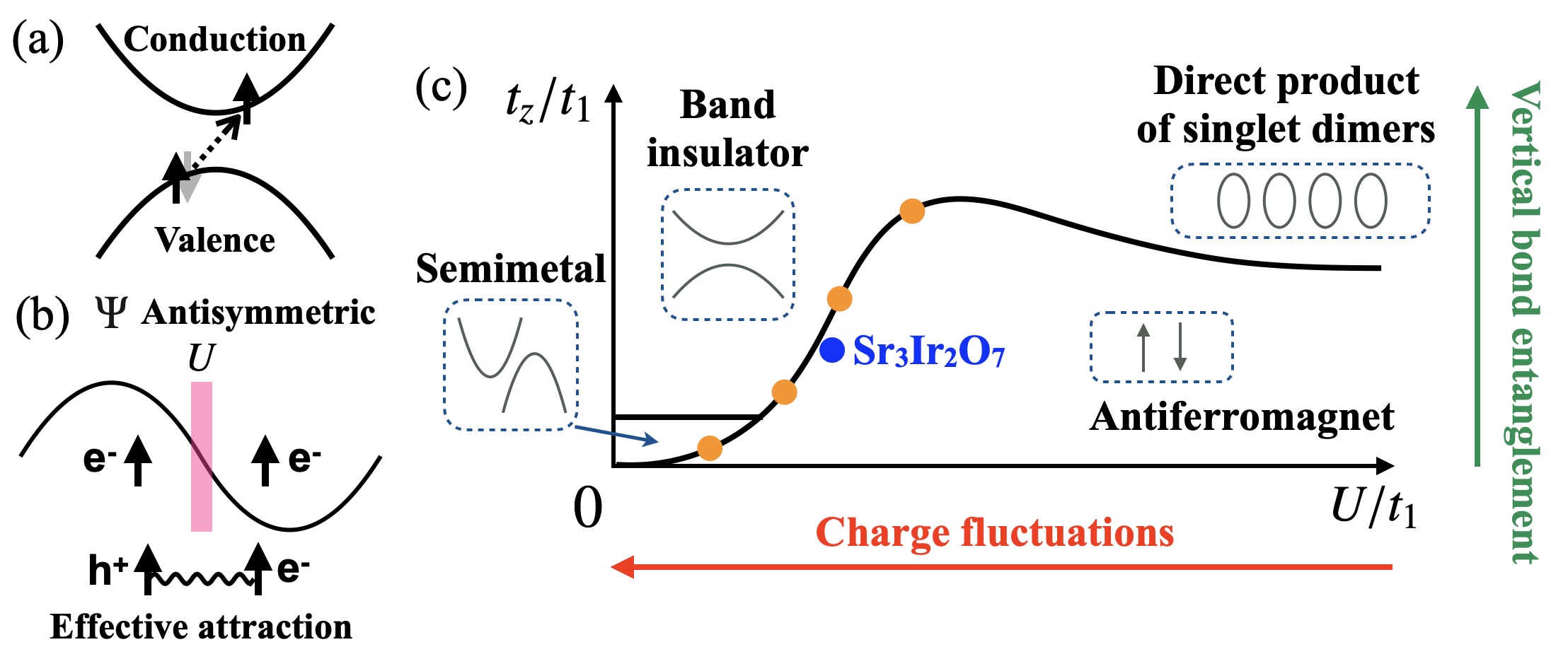}
    \caption{(a)~Schematic picture of electron excitation from a valence band to a conduction band. 
    The spin of the excited electron is flipped, forming a triplet state with the unpaired electron in the valence band. 
    (b)~Wave function of the two-electron problem. 
    Because the spins of the triplet state are symmetric under the exchange of the two electrons, the spatial part of the wave function has to be antisymmetric and is not affected by the on-site Coulomb interaction. 
    As a result, electrons and holes feel an effectively attractive force in the triplet channel. 
    (c)~Schematic phase diagram of the bilayer system for $\alpha \neq 0$ as a function of $U/t_1$ and $t_z/t_1$. 
    The small-$U/t_1$ and $t_z/t_1$ region can be a semimetal, depending on the value of $t_2$.
    The solid circles show the parameter sets for Sr$_3$Ir$_2$O$_7$ and Figs.~\ref{fig-2} and \ref{fig-3}, where $U/t_1=1,2,3$ and $4$.
    }
    \label{fig-1}
\end{figure}
\begin{figure}
    \centering
    \includegraphics[width=\columnwidth]{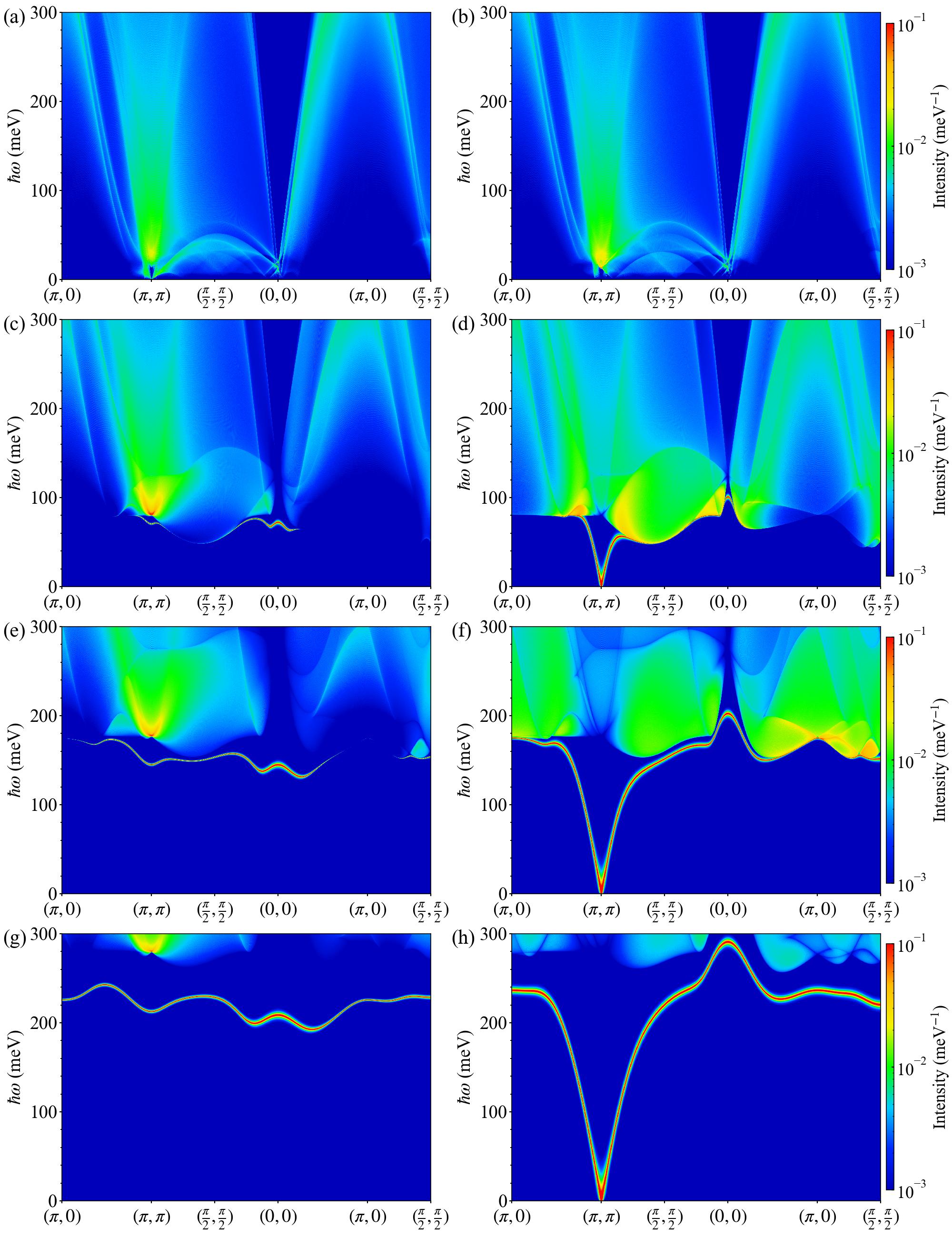}
    \caption{Dynamical spin structure factors of the out-of-phase mode: (a),(c),(e),(g)~in-plane or transverse component $S^{xx}({\bm q}, \omega) = S^{yy}({\bm q}, \omega)$ and (b),(d),(f),(h)~out of plane or longitudinal component $S^{zz}({\bm q}, \omega)$ for (a),(b) $U/t_1=1$; (c),(d) $U/t_1=2$; (e),(f) $U/t_1=3$; and (g),(h) $U/t_1=4$ on the phase boundary between the antiferromagnetically ordered phase and the disordered phase.
    We set the number of sites to $N=2^{21}$ and confirmed the convergence with respect to the system size.}
    \label{fig-2}
\end{figure}
\section{Triplet exciton condensation}
\label{sec:tec}
\vspace{-2mm}
We here discuss the mechanism of triplet exciton formation and the associated BCS-BEC crossover phenomenon.
For $\alpha=0$ in Eq.~\eqref{eq:Hk}, the system is a semimetal in the noninteracting limit.
An infinitesimal $U$ induces the antiferromagnetic (AFM) order due to the perfect Fermi surface nesting~\cite{Golor2014}.
The electron-hole binding and its condensation occur simultaneously at the ordering wave vector $\bm{Q}=(\pi,\pi,\pi)$.
This condensation mechanism is analogous to BCS physics.

For $\alpha \neq 0$, on the other hand, the system can be a band insulator in the noninteracting limit because the spin-orbit coupling allows interband hybridization and creates a narrow charge gap.
A finite Hubbard repulsion $U$ then creates excitonic states below the gap in the triplet channel.
Let us consider an electron excitation with spin-flip from a valence band to a conduction band [Fig.~\ref{fig-1}(a)].
The excited state can be described by a two-body electron-hole problem.
Because the spins of the triplet state are symmetric under  the exchange of electrons, the spatial part of the wave function has to be antisymmetric [Fig.~\ref{fig-1}(b)].
Thus, the energy of the triplet state is practically not affected by the on-site Coulomb repulsion, while the energy of the singlet state increases.
As a result, electrons and holes effectively feel attractive forces in the triplet channel, and the excitonic triplet states emerge below the particle-hole continuum in the presence of a finite $U$~\cite{Suwa2021}.
The exciton binding energy increases with $U$, and the exciton condensation occurs when the binding energy becomes equal to the charge gap, producing an AFM excitonic insulator~\cite{Mazzone2022}.

In the large-$U$ limit, the AFM order can be destroyed by the vertical bond entanglement between the two layers for a large $t_z$.
In the disordered phase, the lowest energy excitation is described by triplons, which represent the triplet state of the two $S=1/2$ spins on a vertical bond~\cite{Lohofer2015,Morreti2015}.
Thus, the electron-hole binding occurs locally in real space.
As a result, the bound state formation and its condensation occur at different energy scales, which is the  characteristic of the BEC regime: the binding occurs at a higher temperature corresponding to an energy scale of order $U$, while the condensation occurs at a lower temperature of order $t_1^2/U$ typically away from the QCP.

The two limiting cases are connected through the BCS-BEC crossover~\cite{Phan2010,Seki2011}.
In the crossover regime, that is, in the intermediate coupling region ($U/t_1 \sim 4$), the 
condensation energy scale reaches its maximum value.
Consequently, the phase boundary between the AFM ordered phase and the disordered phase shows nonmonotonic behavior as a function of $t_z/t_1$ and $U/t_1$ [Fig.~\ref{fig-1}(c)].
This nonmonotonic behavior is in sharp contrast to the monotonic phase boundary for $\alpha = 0$~\cite{Golor2014}, showing the significance of the parameter $\alpha$ and the associated staggered octahedral rotation.

\section{Results}
\label{sec:res}
\vspace{-2mm}
We calculate the dynamical spin structure factor on the phase boundary at zero temperature:
\begin{equation}
S^{\mu\nu}(\boldsymbol{q},\omega)=\int_{\infty}^{\infty}dte^{it\omega}\langle S_{\boldsymbol{q}}^{\mu}(t)S_{-\boldsymbol{q}}^{\nu}(0)\rangle,
\end{equation}
where $S_{\boldsymbol{q}}^{\mu}=\frac{1}{\sqrt{N}}\sum_{{\bm r}}S_{\bm r}^{\mu}e^{i\boldsymbol{q}\cdot\boldsymbol{r}}$ and $N$ is the number of sites, obtained from the dynamical spin susceptibility given in 
Eq.~\eqref{eq:rpa_L}.
The transition points are $(t_z/t_1)_{\rm c} \simeq 0.0856$, $0.5217$, $1.1464$, and $1.8116$, for $U/t_1=1$, $2$, $3$, and $4$, respectively.
Note that in the large-$U$ limit, the critical ratio of the interlayer coupling to the intralayer coupling is $(J_z/J)_{\rm c} \simeq 2.5221$ in the SU(2) symmetric case~\cite{Sen2015}, equivalent to $(t_z/t_1)_{\rm c} \simeq 1.5881$.
Figure~\ref{fig-2} shows the out-of-phase ($q_z=\pi$) transverse response, $S^{xx}(\boldsymbol{q},\omega)=S^{yy}(\boldsymbol{q},\omega)$, and the out-of-phase longitudinal response, $S^{zz}(\boldsymbol{q},\omega)$, for each $U/t_1$. 
The in-phase ($q_z=0$) responses are not shown because the in-phase transverse mode is practically identical to the out-of-phase transverse response~\cite{Kim2012} due to the strong easy-axis anisotropy of the interlayer bond, and the in-phase longitudinal response shows no quasiparticle excitation.
We set $t_1=0.12$ eV, $t_2/t_1=0.1$, and $\alpha=1.4$ typical for iridates~\cite{Mazzone2022}.
The slight difference from the parameter set used in Ref.~\cite{Suwa2021} is irrelevant to the present results because the primary parameter is $U/t_1$.
For $U/t_1=1$, the system transitions from the semimetalic phase to the AFM ordered phase.
Practically no quasiparticle dispersion appears at the transition point, consistent with the BCS regime [Fig.~\ref{fig-2}(a),(b)].
For $U/t_1=2$, clear quasiparticle dispersions appear around ${\bm q}=(0,0)$ and $(\pi,\pi)$, but decay into the particle-hole continuum away from these wave vectors, indicating the crossover nature [Fig.~\ref{fig-2}(c),(d)].
The exciton condensation of the $S^z=0$ state is evidenced by the strong intensity and the linear dispersion around $\bm{q}=(\pi,\pi)$ in the longitudinal mode [Fig.~\ref{fig-2}(d),(f),(h)].
For $U/t_1=3$, close to the value for Sr$_3$Ir$_2$O$_7$, the magnetic energy scale and the charge gap are comparable, and the quasiparticle dispersions appear except near $\bm{q}=(\pi,0)$ [Fig.~\ref{fig-2}(e),(f)].
For $U/t_1=4$, the quasiparticle peaks are well defined over the whole Brillouin zone, implying the onset of the BEC regime [Fig.~\ref{fig-2}(g),(h)].

\begin{figure}
    \centering
    \includegraphics[width=\columnwidth]{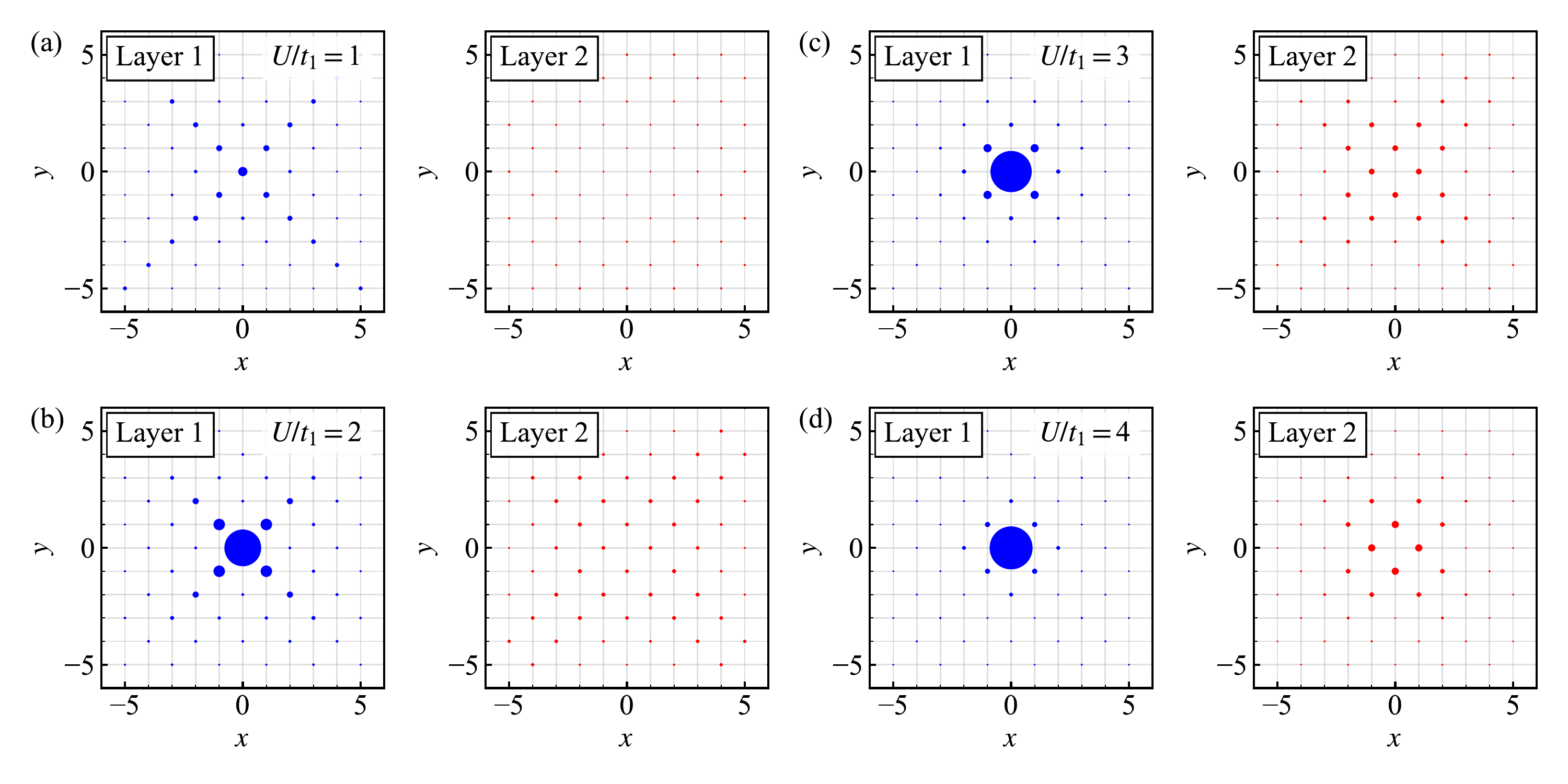}
    \caption{Probability distribution of an electron in the bound state at the transition point for (a) $U/t_1=1$, (b) $U/t_1=2$, (c) $U/t_1=3$, and (d) $U/t_1=4$ given that
    a hole is assumed to be located at the origin of layer 1.
    The area of the circles at each lattice point is proportional to the modulus squared of the normalized exciton wave function.
    $x$ and $y$ represent the two orthogonal directions in each layer in units of the lattice spacing.}
    \label{fig-3}
\end{figure}

Figure~\ref{fig-3} shows the modulus squared of the $S^z=0$ exciton wave function on the phase boundary~\cite{Suwa2021}.
The wave function is extended in real space for $U/t_1=1$, consistent with a very long coherence length (linear size of the particle-hole pair) characteristic of the BCS regime  [Fig.~\ref{fig-3}(a)].
With increasing $U/t_1$, the wave function develops its local nature [Fig.~\ref{fig-3}(b),(c)], and the linear size of excitons eventually becomes $O(1)$ lattice spacing for $U/t_1=4$ [Fig.~\ref{fig-3}(d)], as in the BEC regime.
The amplitude of the quasiparticle wave function depends on the binding energy, which is given by the difference between the lower edge of the electron-hole continuum and the eigenenergy of the quasiparticle state~\cite{Suwa2021}. 
When the binding energy becomes equal to zero, the quasiparticle state disspappers beause the particle and the hole no longer form a bound state.
Thus, with decreasing $U/t_1$, the exciton size increases because the exciton bound state and the associated finite amplitude appear only in restricted wave vectors, as shown in Fig.~\ref{fig-2}.

\section{Conclusion}
\label{sec:con}
\vspace{-2mm}
We argue the mechanism of triplet exciton formation and study the BCS-BEC crossover phenomenon of exciton condensation in a bilayer Hubbard system, calculating the dynamical spin structure factor and the exciton wave function.
On the phase boundary between the disordered phase and the AFM ordered phase, the behavior of exciton condensation significantly depends on the value of $U/t_1$.
In the BCS regime, the quasiparticle dispersion is practically absent, and the exciton wave function is extended in real space.
In the BEC regime, on the other hand, the exciton becomes a local object in real space, and the corresponding quasiparticle has a well-defined dispersion appearing over the whole Brillouin zone.
In contrast to the standard square lattice Hubbard model with real hopping amplitudes, the longitudinal mode appears due to the bilayer structure and the proximity to the QCP.
The crossover regime provides a promising route to functional quantum materials that exploit the rich entanglement between charge, spin, and orbital degrees of freedom.
It is of great interest to study the coupling effect of multiple degrees of freedom and explore intermediate coupling materials in the crossover regime in the near future.

\vspace{-1.5mm}
\section*{Acknowledgments}
\vspace{-3.0mm}
H.S. acknowledges Inamori Research Grants from Inamori Foundation and support from JSPS KAKENHI Grant No. JP22K03508.

\end{document}